\documentclass[12pt,preprint]{aastex}

\newcommand{\solmass} {\hbox{$M_{\odot}$}}
\def\deg{\ifmmode^\circ\else$^\circ$\fi}

\shorttitle{MCP II. The Distance to UGC 3789}
\shortauthors{Braatz et al.}

\begin{document}

\title{The Megamaser Cosmology Project. II. The Angular-Diameter Distance to UGC 3789}


\author{J.A. Braatz}
\affil{National Radio Astronomy Observatory, 520 Edgemont Rd., Charlottesville, VA 22903, USA}
\email{jbraatz@nrao.edu}

\author{M.J. Reid and E.M.L. Humphreys}
\affil{Harvard-Smithsonian Center for Astrophysics, 60 Garden Street, Cambridge, MA 02138, USA}

\author{C. Henkel}
\affil{Max-Planck-Institut f\"{u}r Radioastronomie, Auf dem H\"{u}gel 69, D-53121 Bonn, Germany}

\and

\author{J.J. Condon and K.Y. Lo}
\affil{National Radio Astronomy Observatory, 520 Edgemont Rd., Charlottesville, VA 22903, USA}

\begin{abstract}
The Megamaser Cosmology Project (MCP) aims to determine $H_0$ by measuring angular-diameter distances
to galaxies in the Hubble flow using observations of water vapor megamasers in the circumnuclear accretion disks of active
galaxies.  The technique is based only on geometry and determines $H_0$ in one step, independent of standard
candles and the extragalactic distance ladder.
In Paper I we presented a VLBI map of the maser emission from the Seyfert 2 galaxy UGC 3789.
The map reveals an edge-on, sub-parsec disk in Keplerian rotation, analogous to the megamaser disk in NGC 4258.
Here we present 3.2 years of monthly GBT observations of the megamaser disk in UGC 3789.  We use these observations to
measure the centripetal accelerations of both the systemic and high-velocity maser
components.  The measured accelerations suggest that maser emission lines near the systemic velocity originate on the
front side of the accretion disk, primarily from segments of two narrow rings.  Adopting a two-ring model for the systemic features,
we determine the angular-diameter distance to UGC 3789 to be 49.9 $\pm$ 7.0 Mpc.
This is the most accurate geometric distance yet obtained to a galaxy in the Hubble flow.  
Based on this distance, we determine $H_0$ = 69 $\pm$ 11 km s$^{-1}$ Mpc$^{-1}$.  We also measure the mass of the central black hole
to be 1.09 $\times$ 10$^7$ $\solmass$ $\pm 14\%$.  With additional observations the uncertainty
in the distance to this galaxy can be reduced to under 10\%.  Observations of megamaser disks in 
other galaxies will further reduce the uncertainty in $H_0$ as measured by the MCP.
\end{abstract}


\section{Introduction}

As a complement to observations of the cosmic microwave background (CMB), an independent
measurement of the Hubble Constant, $H_0$, within 3\% rms would place a valuable 
constraint on the dark energy equation-of-state parameter, $w$ (e.g. Hu 2005, Olling 2007).  
The most widely accepted value
for $H_0$ is 72 $\pm$ 7 km s$^{-1}$ Mpc$^{-1}$ (Freedman et al. 2001), based on the ``extragalactic
distance ladder'' and using the Large Magellanic Cloud (LMC) to calibrate Cepheid variables which
are treated as standard candles.  The Cepheid metallicity correction has been controversial,
as highlighted by Sandage et al. (2006), who used similar methods but a different metallicity
correction to obtain $H_0$ = 62 $\pm$ 6 km s$^{-1}$ Mpc$^{-1}$.  If measured at the 
$\lesssim$ 3\% level, a value of $H_0$ = 72 km s$^{-1}$ Mpc$^{-1}$ would be consistent with 
the cosmological constant as an explanation of dark energy ($w$ = -1), whereas 
$H_0$ = 62 km s$^{-1}$ Mpc$^{-1}$ would challenge that model (e.g. Figure 14 of 
Riess et al., 2009).

Macri et al. (2006) used NGC 4258 rather than the LMC to calibrate Cepheids for the extragalactic
distance scale and measured $H_0$ = 74 $\pm$ 7 km s$^{-1}$ Mpc$^{-1}$.  Riess et al. (2009) subsequently
refined the measurement, also using Cepheids in NGC 4258 to calibrate those in selected SN Ia host galaxies, 
and they determined $H_0$ = 74.2 $\pm$ 3.6 km s$^{-1}$ Mpc$^{-1}$ (5\%).  Riess et al. are able 
to reduce the systematic uncertainty in $H_0$ compared to the LMC-based results because the Cepheids in 
NGC 4258 are more similar in metallicity and period to those in the galaxies being calibrated.  NGC 4258 
is a good foundation for the 
extragalactic distance scale because its distance (7.2 Mpc $\pm$ 7\%) is known to high accuracy 
from observations of circumnuclear water megamasers in its AGN accretion disk (Herrnstein et al. 1999).
NGC 4258 is too close, though, for determining $H_0$ directly because its recession velocity may be
highly influenced by its peculiar motion.  Lo (2005) gives a review of megamasers,
including a discussion of the maser distance technique used in NGC 4258.

The Megamaser Cosmology Project (MCP) aims to determine $H_0$ by measuring angular-diameter distances to galaxies 
in the Hubble flow, $\sim$ 50 -- 200 Mpc distant, using the maser technique.  Being independent of
Cepheids and the CMB, the MCP will provide a valuable check of these other methods for measuring $H_0$.
The project is a multi-year effort that begins with surveys for new megamaser disks.  Measuring a
distance to each galaxy detected with a megamaser disk requires high-fidelity VLBI imaging, which we pursue with the 
High Sensitivity Array (VLBA+GBT+VLA+Effelsberg), and spectral-line monitoring to measure the centripetal
accelerations of individual lines in the maser spectrum.  The MCP aims to measure maser distances with $\sim$ 10\% 
or better accuracy to each of about 10 galaxies, thereby determining H$_0$ to $\sim$ 3\%.

In a survey with the Green Bank Telescope, Braatz \& Gugliucci (2008; hereafter BG08) discovered water maser 
emission from the Seyfert 2 galaxy UGC 3789.  The maser spectrum has three sets of Doppler
components spaced nearly symmetrically at and around the galaxy's systemic recession velocity (Figure 1). This 
profile is characteristic of megamasers in an edge-on accretion disk.  The megamaser in UGC 3789 is 
similar to that in NGC~4258, where ``high-velocity'' blue-shifted and red-shifted maser components arise from 
the tangent points of the disk rotating toward and away from the observer, 
and components near the systemic recession velocity
originate on the near side of the disk, along the line of sight to the central black hole.  

Rotation velocities indicated by the high-velocity maser lines in UGC 3789 are $\sim$ 400--800 km s$^{-1}$.  By 
monitoring the maser spectrum roughly monthly over 6 months, BG08 measured secular, redward drifts in radial velocity
of 2 -- 8 km s$^{-1}$ yr$^{-1}$ among the systemic group of maser components, while the red- and blue-shifted 
features remained relatively fixed in velocity.  The radial velocity drift is interpreted as centripetal 
acceleration in the maser gas as it orbits the central supermassive black hole.  The spectral profile and
the measured velocity drifts demonstrate that the maser disk in UGC 3789 is suitable for
measuring an angular-diameter distance to the galaxy, using the method pioneered by Herrnstein et al. (1999).  

In Paper I from the Megamaser Cosmology Project (Reid et al. 2009), we presented a VLBI image
of the maser disk in UGC 3789.  The masers reveal nuclear gas in an edge-on disk ranging in radii 
from 0.08 to 0.30 pc.  The high-velocity lines trace a Keplerian rotation curve with high precision.  
While the high-velocity lines indicate relatively little disk warping, the systemic features show some structure 
that suggests the front side of the disk may be warped, tilted, or geometrically thick.

Here we present single-dish observations of the UGC 3789 maser made with the Green Bank
Telescope over a period of 3.2 years.  Using these observations we refine the acceleration
measurements made by BG08 and we present a model to estimate the distance to UGC 3789 using the
maser observations.

\section{Observations}
Between 2006 January 12 and 2009 March 31, we observed the 22 GHz water maser emission from
UGC 3789 using the Green Bank Telescope (GBT) of the National Radio Astronomy Observatory.  
We observed the source 36 times during this period, with the observations spaced roughly monthly 
throughout, except for the summer months when the humidity makes observations at 22 GHz inefficient.
Table 1 shows the observing date and sensitivity for each observation.  The 
first nine epochs are the same as reported in BG08.

We configured the spectrometer with two 200 MHz spectral windows, 
each with 8192 channels, giving a channel spacing of 24 kHz, corresponding to
0.33 km s$^{-1}$ at 22 GHz.  The first spectral window was centered on the
systemic recession velocity of the galaxy, and the second was offset by 180 MHz
to the red, giving a total bandwidth coverage of 380 MHz (5100 km s$^{-1}$). 
We observed both circular polarizations simultaneously using a total-power nodding mode in which the galaxy was 
placed alternately in one of the two beams of the 18 -- 22 GHz K-band receiver, observing 
for 2.5 minutes at each position and then repeating the cycle.

We reduced and analyzed the data using GBTIDL.  In order to enhance weak 
signals, we smoothed the blank sky reference spectra using a 16-channel boxcar function
prior to calibration.  We corrected the spectra for atmospheric 
attenuation using opacities derived from weather data, and applied an 
elevation-dependent gain curve.  Our flux density scale is uncertain by about 15\%, 
limited by noise-tube calibration and pointing errors.  To remove residual baseline shapes 
from each 200 MHz spectrum, we subtracted a polynomial (typically third order) fit to the 
line-free channels of the calibrated spectrum.  Finally, we Hanning smoothed the spectra.
Integration times were typically about 2 hours for each epoch prior to May 2007, and about 4 hours after.

Unless stated otherwise, velocities quoted in this paper use the optical convention (v$_{opt}$ = c$z$) and
are given in the frame of the local standard of rest (LSR).  For UGC 3789, the LSR frame relates to the
heliocentric frame by $v_{hel} = v_{LSR}$ - 0.3 km s$^{-1}$.  

\section{Accelerations in UGC 3789}

In Figure 2 we plot the radial velocities of UGC 3789 maser peaks as a function of the
epoch observed.  The local maxima in the maser profiles were determined by eye and are accurate
to about one channel width (0.3 km s$^{-1}$).  The trends are evident: high-velocity spectral lines are roughly invariant in velocity
throughout the monitoring period, while the radial velocities of the systemic features show a clear drift redward.
The maser spectral lines in the systemic range remain within a window of velocities such that new lines may
appear near the low velocity limit and disappear beyond the high velocity limit.  Some lines can be tracked 
throughout the entire monitoring period, while others appear or disappear over the course of the monitoring.  
If sufficiently long-lived, it would take individual maser spectral components at least 10 years, and in some 
cases much longer, to drift through the entire velocity window.

The drift pattern evident in Figure 2 is consistent with the picture of high-velocity maser emission originating near 
the midline of the edge-on megamaser disk, and maser emission near the systemic velocity originating on the near 
side of the disk.  Here we define the midline as the diameter of the disk perpendicular to the line of sight.
Maser features along the 
line of sight to the central black hole but on the far side of the disk would show a drift to lower velocities, 
but there is no evidence in our data for such blueward drifts in UGC 3789.  The absence of detected masers from
the far side of the disk may be understood if systemic masers amplify radio continuum emission originating interior 
to the maser disk.  In that case masers on the far side of the disk would be beamed away from us.  Alternatively, 
maser emission from the far side of the disk might be free-free absorbed near the central black hole.

To measure accelerations for each of the maser lines in UGC 3789, we applied a global, nonlinear, 
least-squares $\chi^2$ minimization routine that decomposes the spectra into multiple Gaussians at 
each epoch and solves simultaneously for the amplitude, width, velocity, and constant drift
(acceleration) for each maser component.  The method is described by Humphreys et al. (2008).
We allowed the amplitude, width, and velocity to vary for each spectral component from epoch to epoch.

Finding consistent and stable solutions from the fitting algorithm required experimentation with the 
initial parameters used by the fitting algorithm, and many trials were required to determine the 
final accelerations.  We found it convenient to divide the spectra into 15 -- 30 km s$^{-1}$ 
segments, selected by choosing edges near local minima in the profile.  We then solved for accelerations in each
segment separately.  We began the fitting process using initial parameters describing
equally spaced spectral components across the velocity range of interest.  In the initial trials we spaced spectral
components by about 4 km s$^{-1}$.  We then added components in later trials until the solution was satisfactory.  

After each trial, we judged the results by examining the fits and residuals at each epoch.
Figure 3 gives an example of the Gaussian components and residuals determined from a good fit to a section 
of the data among the systemic components.  We also compared the fitted accelerations to the trends 
apparent in plots showing velocities of maser peaks over time, as in Figure 2.  The line segments plotted
in the central panel of Figure 2 correspond to accelerations measured from the global fit, while the plus symbols 
represent the velocities of maser peaks 
determined by eye from each spectrum.  Where the fit accelerations were clearly in disagreement with the
trends in velocities of the maser peaks, we rejected the solution and tried a new set of initial parameters.
When the residuals were poor over the entire velocity range being fitted, we added components and
maintained equal velocity spacing.  When the residuals were poor only in a narrow range of the velocities being
fit, we kept a base set of equally-spaced spectral components and inserted additional spectral components
at the specific velocities where the residuals were unsatisfactory.  A good fit to masers in the systemic 
velocity range typically required that we use one Gaussian component for every 2 -- 3 km s$^{-1}$ of velocity space
being fit.
We confirmed that the solutions remained stable and consistent near the edges of each segment by running test 
cases with overlap on the segment boundaries.

It is apparent in Figure 2 that the fitting algorithm is able to find stable solutions even in parts of
the spectrum where there are no distinct peaks in the spectral profile.  In these velocity ranges the 
emission lines are typically weak or 
heavily blended.  We note, however, that the data used in the final distance analysis come only from
velocity ranges where there are both distinct peaks detected by eye, and stable solutions determined from
the fitting algorithm.

For high-velocity lines we used data from all 36 epochs to fit for acceleration.  The small accelerations 
among high-velocity lines make it possible to track lines for long durations and obtain stable solutions 
using data for all 3.2 years of monitoring.  For the systemic features, however, we found that using 
all 36 epochs made the global solution unstable.  Fitting systemic lines using all epochs may be complicated 
by the appearance and disappearance of many fast-drifting spectral components, making it difficult to account 
for all components over long durations.  Our aim is to determine accelerations appropriate
for the VLBI observing date of 2006 December 10.  We therefore fit accelerations of systemic components 
using a subset of the epochs roughly centered on this date, and including at least the epochs between 
09 September 2006 and 14 April 2007.  In Figure 2, the lengths of the line segments showing the acceleration
fits represent the number of epochs used in each fit.

In Figure 4 we show the best-fit accelerations for each maser line as measured by the global least-squares fit.
The redshifted lines have a mean value of -0.13 $\pm$ 0.04 km s$^{-1}$ yr$^{-1}$ and the blueshifted 
lines have a mean of -0.16 $\pm$ 0.04 km s$^{-1}$ yr$^{-1}$.  Here we quote unweighted, standard
errors of the mean.  Accelerations of the 
systemic features are consistent with the preliminary work of BG08 and range from 
0.5 to 7.3 km s$^{-1}$ yr$^{-1}$.  Tables 2 -- 4 list the accelerations 
measured for each line.  The uncertainties listed in these tables are formal values from the
global fit, and may underestimate the true uncertainties.

\section{Modeling Systemic Maser Emission in a Thin Ring}

In this section we discuss the ``megamaser disk method'' for measuring the angular-diameter 
distance to a galaxy based on a straightforward model of the maser emission arising from an edge-on
Keplerian disk.  This model gives a good approximation to the maser emission from NGC 4258.  To measure the 
distance to UGC 3789, we make a simple extension of this model, as presented in Section 5.

First we consider a model in which maser emission originates in a thin, flat, edge-on disk and the dynamics
are dominated by a central massive object, with all maser clouds in circular orbits.  In this model,
high-velocity masers trace gas near the disk midline.  Systemic
masers occupy part of a ring orbiting at a single radius, and covering a small range of azimuthal angles on the 
near side of the disk.  When plotted on a position-velocity (P-V) diagram, the high-velocity masers trace a Keplerian 
rotation curve and the systemic masers fall on a line segment.  If the line segment is extended 
to the rotation curve traced by the high-velocity features (as in the dotted lines shown in Figure 3 of 
Paper I), the intersecting points ($\theta_1,v_1$) and ($\theta_2,v_2$) determine both the angular 
radius ($\theta$) and magnitude of the rotation velocity ($v_r$) of the narrow
ring traced by systemic masers, according to $\theta = \left |\theta_1 - \theta_2 \right |/2$ and 
$v_r = \left |v_1 - v_2 \right |/2$.  In this model, the angular-diameter distance to the galaxy is 
\begin{equation}
D_a = r/\theta = v_r^2/(a \theta),
\end{equation}
where $r$ is the radius of the ring traced by the systemic maser components
($r = v_r^2/a$) and $a$ is the centripetal acceleration of systemic components, measured from the change in 
Doppler velocity with time.

In practice, we do not measure $\theta$ and $v_r$ directly.  Instead we measure the Keplerian
rotation constant $k$, which we determine by fitting the high-velocity lines to the
rotation curve $v = k (\theta - \theta_0)^{-1/2} + v_{sys}$.  In addition, we measure a curvature parameter
$\Omega$, which is the slope of the line traced by systemic features on the P-V diagram ($\Omega \simeq dv/d\theta$
for the systemic features).  With this parametrization, we can write the angular-diameter distance to the galaxy as 
\begin{equation}
D_a = a^{-1} k^{2/3} \Omega^{4/3}. 
\end{equation}
The fractional uncertainty in $D_a$ is 
\begin{equation}
\frac{\sigma_{Da}}{D_a} \simeq \sqrt{\left ( \frac{\sigma_a}{a} \right )^2 + \frac{4}{9} \left (
\frac{\sigma_k}{k} \right )^2 + \frac{16}{9} \left (\frac{\sigma_\Omega}{\Omega} \right )^2}.
\end{equation}

Here we use a notation in which the uncertainty in a measured or derived value, $x$, is represented as $\sigma_x$, so
the fractional uncertainty in $x$ is $\frac{\sigma_x}{x}$.

The straightforward method for measuring the angular-diameter distance described above works well for NGC 4258, where
systemic accelerations fall in the range 7.7 -- 8.9 km s$^{-1}$ yr$^{-1}$ (Humphreys et al 2008), 
indicating they are seen within a narrow range of radii, $r$, from the central black hole 
($r \propto$ 1/$\sqrt{a}$).  The tightly confined range of radii in NGC 4258 is a consequence 
of the masers forming in the bottom of a bowl-shaped region of the warped accretion disk 
(Humphreys et al. 2008).  While it may be fortuitous that the systemic masers prefer a single
radius, we note that the high-velocity features in NGC 4258 also seem to prefer specific radii, 
and even show periodicity suggestive of spiral structure (Maoz 1995, Humphreys et al. 2008).

The masers in UGC 3789 also appear to cluster around discrete radii from the black hole.
In Figure 5 we reproduce the VLBI map presented in Paper I.  The
``red'' features in particular are seen mainly at three distinct radii from the central black hole. 
Evidence that the systemic features are also clustered in radius comes from the acceleration profile
(Figure 4), as we describe in the next section.  

\section{Fitting UGC 3789 with the Thin Ring Model}

Except for the westernmost ``red'' maser, all the high-velocity maser lines in UGC 3789 fall on a 
straight line, within the measurement uncertainties (Figure 5).  The line has a position angle 
of 41.1$\deg$, east of north.  The maser distribution is consistent with emission from
a thin, edge-on disk.  
The systemic maser spots show systematic deviations from the line defined by the high-velocity masers, 
with the easternmost systemic masers appearing north of the line and the westernmost systemic masers 
south of it.  One way to interpret this arrangement is that the near side of the disk, traced
by the systemic masers, is warped while the high-velocity lines trace a part of the
disk that appears relatively unwarped from our point of view.

We define an ``impact parameter'' for each maser spot as the projection of its angular offset
onto the line defined by the high-velocity masers, with an impact parameter $\theta_0$ identifying the 
location of the dynamical center. To determine the impact parameters, we rotated the map on the 
sky by 48.9$\deg$ (counterclockwise, when viewed as in Figure 5) and measured the impact parameter for 
each maser component by its abscissa on the rotated axes.

When modeling the data, we corrected the recorded ``optical'' velocities of all maser spots 
for relativistic effects, as in Argon et al. (2007).  First we recalculated each velocity using the 
relativistic Doppler formula.  We estimated the black hole mass using these velocities, then corrected for the 
gravitational time dilation caused by the central black hole.  Finally we corrected velocities
of the systemic features for the relativistic transverse Doppler effect.  The effect of the relativistic
corrections is to reduce the measured optical velocities by 12 -- 27 km s$^{-1}$, the blueshifted features 
getting the smaller corrections.  We then used these corrected velocities in all of the analysis, including 
the determination of the final black hole mass reported in this paper.

\subsection{The Keplerian Rotation Curve}

As a matter of convenience in fitting for the Keplerian rotation constant $k$, we write the impact 
parameter $\theta$ in terms of the 
independent variable, $v$: $\theta(v) = k^2(v-v_0)^{-2} + \theta_0$.  Uncertainties in the velocities are 
negligible compared to those in the impact parameters and are ignored in this analysis.  We find the best 
fit has $\chi^2$ per degree of freedom of 8.5, so we scale the $k^2$ uncertainty by $\sqrt{8.5}$.  
We therefore determine $k = 440.4 \pm 1.3$ km s$^{-1}$ mas$^{0.5}$.  The best-fit value for the recession 
velocity from the fit to the Keplerian rotation curve is 3270 $\pm$ 4 km s$^{-1}$ (relativistic velocity).  
The apparently large $\chi^2$ per degree of freedom results from small-amplitude perturbations in the positions 
and velocities of maser components, as might be expected, for example, with position-velocity perturbations 
from spiral density waves.  In this analysis, the position of the black hole along the disk plane ($\theta_0$)
is treated as a free parameter.  If we were to fix $\theta_0$ to the centroid of the systemic
maser spots, 0.04 mas from the best-fit position, we would measure $k = 445.4$ km s$^{-1}$ mas$^{0.5}$.  
The difference in $k$ is insignificant in the calculation of distance, but it suggests that the formal 
uncertainty in $k$ may underestimate the true value.  To be conservative, we use 
$k = 440.4 \pm 3.0$ km s$^{-1}$ mas$^{0.5}$ in our calculations, noting that $k$ is still a negligible 
contributor to the total error in the distance.

\subsection{Accelerations as Evidence for a Two Ring Model}

The wide range of systemic accelerations measured in UGC 3789 (0.5 -- 7.3 km s$^{-1}$ yr$^{-1}$)
indicates that the systemic maser emission is observed over a larger range of radial distance from 
the black hole than in NGC 4258.  Accelerations on the low-velocity end of the systemic spectrum 
are primarily in the range 1 -- 2 km s$^{-1}$ yr$^{-1}$, while those on the high-velocity end are
primarily 4 -- 6 km s$^{-1}$ yr$^{-1}$.  This distribution suggests that the maser clouds
lie mainly in two rings.  We will identify masers having observed velocities of 3232 $<$ c$z$ $<$ 3271 km s$^{-1}$ 
with ``ring 1'' and those having 3292 $<$ c$z$ $<$ 3318 km s$^{-1}$ with ``ring 2'', where here
we use the optical convention for velocities, as in Table 3.
The mean accelerations are 1.50 $\pm$ 0.13 km s$^{-1}$ yr$^{-1}$ for ring 1, and 4.96 $\pm$ 0.34 
km s$^{-1}$ yr$^{-1}$ for ring 2.  The uncertainties here are unweighted, standard errors of the mean.
Masers with intervening velocities in the range 3271 -- 3292  km s$^{-1}$ have more scattered
accelerations and are not considered in this analysis.
We note that there is a distinct separation even in the single-dish spectrum between masers 
associated with ring 1 and those with ring 2 (Figure 1 lower panel).

In this analysis, we include the approximation that all the high-velocity 
masers fall on the midline, so individual accelerations of high-velocity features do not enter 
into the distance calculation directly.  Although the high-velocity lines have accelerations near
zero, the tendency to negative values suggests that they form slightly on the back side 
of the disk midline. The uncertainty in the distance to UGC 3789 introduced by this deviation 
is not significant compared to our measurement uncertainties.

\subsection{Curvature Parameter $\Omega$}

Figure 6 shows the distribution of systemic maser features on a P-V diagram.  Rather than tracing 
a single line, as would be expected if all masers were at a common radial distance from the 
central black hole, there is a kink in the distribution with the masers on each side of the kink
roughly defining a line.  The distribution is consistent with the masers being detected primarily 
in two rings.  In Figure 6 we use unique symbols to differentiate the masers associated
with each of the rings.  We can measure the slope $\Omega$ separately for each ring:
$\Omega = 695 \pm 65$ km s$^{-1}$ mas$^{-1}$ for ring 1 and $\Omega = 1654 \pm 440$ 
km s$^{-1}$ mas$^{-1}$ for ring 2.  The poorer fit for ring 2 results, in part, from that ring
covering a narrower velocity range than ring 1, so it has fewer spectral channels to fit
from the VLBI map.

\subsection{The Angular-Diameter Distance to UGC 3789}

We can determine the distance to UGC 3789 separately for each of the two rings using Equation 2.  
For ring 1 we find $D_a = 50.2 \pm 7.7$ Mpc, and for ring 2 $D_a = 48.1 \pm 17.4$ Mpc.  The weighted
mean of these two values gives our best determination of the maser distance to UGC 3789: $D_a$ = 49.9 $\pm$ 7.0 Mpc
(14\%).  

The rotation curve determined from high-velocity lines provides a central black hole mass 
$M_{bh} = 1.13 \times (\frac {k}{km\,s^{-1}\,mas^{0.5}})^2 \times (\frac {D_a}{Mpc})\ \solmass$.
At 49.9 Mpc and $k$ = 440 km s$^{-1}$ mas$^{0.5}$, the mass of the nuclear black hole is 
1.09 $\times$ 10$^7 \solmass \pm 14\%$, with nearly all of the mass uncertainty being caused by the distance uncertainty.  

\section{Discussion}

The peculiar radial velocity of UGC 3789 relative to the CMB is -151 $\pm$ 163 km s$^{-1}$ (Masters, 
private communication), using a flow model based on the SFI++ sample (Masters et al. 2006; 
Springob et al. 2007).  Here peculiar velocity is defined as $v_{pec} = v_{rec} - H_0 D$, where v$_{rec}$ is 
the galaxy's observed recession velocity.  Writing the UGC 3789 maser radial velocity in the CMB frame 
($v_{CMB} \simeq v_{LSR}$ + 60 km s$^{-1}$) and correcting for the peculiar velocity, we obtain 
a relativistic, recessional flow velocity of 3481 $\pm$ 163 km s$^{-1}$.  Applying a standard $\Lambda$CDM
model with $\Omega_m$ = 0.26, we therefore determine $H_0$ = 69 $\pm$ 11 km s$^{-1}$ Mpc$^{-1}$.

The main source of uncertainty in the distance to UGC 3789 comes from the measurement of the
orbital curvature parameter $\Omega$.  The uncertainty in the acceleration is also a significant contributor, 
while the contribution from the Keplerian rotation constant is negligible.  Therefore we can best reduce 
the total uncertainty in the distance to UGC 3789 by obtaining a more sensitive VLBI map.  
We are currently processing additional VLBI observations of UGC 3789, which we can average to increase the
map sensitivity.  Improving the measurement of the acceleration profile requires more sensitive monitoring 
observations.  Recognizing this need, we increased the duration of GBT observations after May 2007.  
These efforts should reduce the overall distance uncertainty to $<$ 10\%, which would determine $H_0$ with
accuracy comparable to the Hubble Key Project, but with entirely independent data and methods.

Ultimately, to reach an uncertainty in $H_0$ of $\lesssim$ 3\% from the Megamaser Cosmology Project, we 
require distances to at least 10 galaxies comparable to UGC 3789.  We are currently pursuing 
observations of the megamaser galaxies Mrk 1419, NGC 6264, and NGC 6323, which could eventually lead to 
comparable distance measurements.  There are another $\sim$ 6 maser disk galaxies known that may also
be suitable for distance measurements, but the quality of these (i.e. brightness of the masers, 
richness of the spectrum, and velocity coverage) is not as good.  Surveys for additional maser galaxies 
therefore remain an important part of the MCP. 

\section {Summary}

As part of the Megamaser Cosmology Project, we have observed the water megamaser disk in the
nucleus of the galaxy UGC 3789.  Paper I (Reid et al. 2009) presents a VLBI map of the maser emission
and here we present measurements of orbital accelerations from GBT monitoring observations spanning
3.2 years.  The 
high-velocity maser emission traces a thin, edge-on disk 0.08 to 0.30 pc in radius, and demonstrates 
Keplerian rotation about a 1.09 $\times$ 10$^7 \solmass$ black hole.  The dynamical center corresponds to
a (relativistic) LSR recession velocity of 3270 $\pm$ 4 km s$^{-1}$.  We model the emission from
the systemic masers using two narrow rings and apply the ``maser distance method'' to each.
Combining the results, we determine a distance to UGC 3789 of 49.9 $\pm$ 7.0 Mpc, corresponding to 
$H_0$ = 69 $\pm$ 11 km s$^{-1}$ Mpc$^{-1}$.  Additional observations of the maser in
UGC 3789 will improve the signal-to-noise ratio and can reduce the distance uncertainty to about 10\%.  
We have demonstrated that angular-diameter distance determination to galaxies with observed megamaser
disks is a promising approach to determining $H_0$, independent of the traditional distance ladder
approach based on Cepheid variables.  The Megamaser Cosmology Project aims to measure the distances 
to additional megamaser disk galaxies in order to reduce the overall uncertainty
in $H_0$ and eventually place meaningful constraints on the dark energy equation-of-state 
parameter.

\acknowledgements
We thank the referee for helpful comments, and we are grateful to the NRAO
staff in Green Bank for their many contributions to this research program.
The National Radio Astronomy Observatory is a facility of the National 
Science Foundation operated under cooperative agreement by Associated
Universities, Inc.

\clearpage

\begin{table}
\small
\caption{Observation dates and sensitivities for each GBT spectrum used to measure accelerations.
The sensitivities are calculated after Hanning smoothing and are based on 0.33 km s$^{-1}$ channels.}
\bigskip
\begin{tabular}{lrrr}
\tableline\tableline
Date & Day number & T$_{sys}$ (K) & RMS sensitivity (mJy) \\
\tableline
2006 Jan 12  &    0 &  44  &   2.3 \\
2006 Feb 08  &   27 &  37  &   1.5 \\
2006 Mar 05  &   53 &  43  &   2.0 \\
2006 Mar 11  &   58 &  61  &   4.2 \\
2006 Mar 26  &   74 &  43  &   1.8 \\
2006 Apr 17  &   96 &  62  &   3.0 \\
2006 May 01  &  110 &  39  &   1.7 \\
2006 May 15  &  123 &  67  &   3.0 \\
2006 May 23  &  132 &  44  &   1.9 \\
2006 Sep 09  &  241 &  83  &   5.3 \\
2006 Sep 25  &  256 &  60  &   2.8 \\
2006 Oct 30  &  291 &  57  &   2.5 \\
2006 Nov 21  &  313 &  36  &   1.2 \\
2006 Nov 28  &  320 &  36  &   1.3 \\
2006 Dec 16  &  338 &  41  &   1.6 \\
2007 Jan 20  &  373 &  37  &   1.5 \\
2007 Feb 05  &  389 &  34  &   1.5 \\
2007 Feb 22  &  406 &  57  &   2.1 \\
2007 Mar 29  &  442 &  49  &   1.7 \\
2007 Apr 14  &  457 &  43  &   1.5 \\
2007 Oct 01  &  627 &  46  &   1.2 \\
2007 Oct 29  &  655 &  33  &   0.94 \\
2007 Nov 28  &  685 &  41  &   0.92 \\
2007 Dec 30  &  717 &  42  &   0.85 \\
2008 Jan 30  &  749 &  34  &   0.75 \\
2008 Feb 28  &  778 &  34  &   0.91 \\
2008 Mar 25  &  804 &  44  &   1.2 \\
2008 Apr 24  &  834 &  56  &   1.6 \\
2008 May 28  &  868 &  52  &   1.4 \\
2008 Sep 20  &  982 &  48  &   1.4 \\
2008 Oct 27  & 1019 &  41  &   1.0 \\
2008 Nov 28  & 1051 &  44  &   1.2 \\
2009 Jan 09  & 1093 &  37  &   0.96 \\
2009 Jan 31  & 1115 &  33  &   1.2 \\
2009 Mar 04  & 1147 &  32  &   0.86 \\
2009 Mar 31  & 1174 &  47  &   1.4 \\
\tableline
\end{tabular}
\end{table}

\clearpage

\begin{table}
\tiny
\caption{Results from the global fitting program for ``blue'' high-velocity features in UGC 3789.  The columns show (1) the 
LSR velocity of the maser component at the reference epoch (2006 December 10) determined from the Gaussian fit, (2) 
the uncertainty in the fit velocity, (3) the full width at half maximum of the Gaussian fit at the reference epoch, 
(4) the uncertainty in the width, (5) the fit acceleration, and (6) the uncertainty in the acceleration.
Velocities use the optical convention.}
\bigskip
\begin{tabular}{lrrrrr}
\tableline\tableline
v$_{LSR}$ & $\sigma_{v}$ & Width & $\sigma_{width}$ & Accel & $\sigma_{accel}$ \\
(km s$^{-1})$ & (km s$^{-1}$) & (km s$^{-1}$) & (km s$^{-1})$ & (km s$^{-1}$ yr$^{-1}$) & (km s$^{-1}$ yr$^{-1}$) \\
\tableline
2468.05 &    0.08 &    2.66 &    0.48 &   -0.69 &    0.06 \\
2472.17 &    0.03 &    2.34 &    0.37 &   -0.48 &    0.05 \\
2474.67 &    0.05 &    2.38 &    0.39 &   -0.36 &    0.05 \\
2477.91 &    0.07 &    2.91 &    0.44 &   -0.26 &    0.06 \\
2544.05 &    0.04 &    3.57 &    0.38 &   -0.04 &    0.04 \\
2550.06 &    0.05 &    3.17 &    0.39 &    0.33 &    0.05 \\
2565.69 &    0.05 &    3.96 &    0.25 &   -0.49 &    0.02 \\
2568.75 &    0.06 &    2.83 &    0.36 &    0.04 &    0.03 \\
2572.66 &    0.03 &    2.74 &    0.39 &    0.06 &    0.03 \\
2577.69 &    0.04 &    2.88 &    0.41 &   -0.16 &    0.04 \\
2585.18 &    0.06 &    2.36 &    0.35 &    0.06 &    0.05 \\
2588.53 &    0.02 &    2.53 &    0.21 &   -0.21 &    0.03 \\
2591.47 &    0.04 &    2.55 &    0.28 &   -0.51 &    0.03 \\
2594.75 &    0.03 &    2.78 &    0.27 &   -0.27 &    0.02 \\
2604.26 &    0.02 &    3.39 &    0.38 &    0.40 &    0.04 \\
2608.58 &    0.04 &    3.08 &    0.33 &    0.37 &    0.03 \\
2612.02 &    0.03 &    2.96 &    0.24 &   -0.47 &    0.03 \\
2615.66 &    0.03 &    3.24 &    0.24 &   -0.73 &    0.03 \\
2629.70 &    0.05 &    3.41 &    0.33 &    0.12 &    0.05 \\
2633.39 &    0.04 &    3.22 &    0.29 &   -0.32 &    0.03 \\
2637.35 &    0.05 &    2.97 &    0.39 &   -0.29 &    0.05 \\
2644.94 &    0.16 &    2.28 &    0.44 &    0.29 &    0.09 \\
2649.86 &    0.06 &    2.34 &    0.41 &    0.52 &    0.05 \\
2655.45 &    0.10 &    2.20 &    0.40 &   -0.02 &    0.08 \\
2659.43 &    0.06 &    2.33 &    0.40 &   -0.51 &    0.05 \\
2667.76 &    0.05 &    3.55 &    0.45 &    0.05 &    0.04 \\
2672.00 &    0.09 &    2.93 &    0.55 &    0.04 &    0.08 \\
2676.38 &    0.05 &    2.40 &    0.31 &   -0.05 &    0.04 \\
2678.72 &    0.07 &    2.22 &    0.32 &    0.10 &    0.04 \\
2680.98 &    0.04 &    2.39 &    0.22 &   -0.01 &    0.02 \\
2683.75 &    0.04 &    2.72 &    0.18 &   -0.18 &    0.03 \\
2686.15 &    0.04 &    2.60 &    0.18 &   -0.13 &    0.04 \\
2688.51 &    0.03 &    2.68 &    0.14 &   -0.29 &    0.02 \\
2691.23 &    0.03 &    3.00 &    0.16 &   -0.53 &    0.03 \\
2694.07 &    0.03 &    3.16 &    0.25 &   -0.31 &    0.03 \\
2697.66 &    0.03 &    4.50 &    0.29 &   -0.19 &    0.03 \\
2701.07 &    0.06 &    2.68 &    0.38 &   -0.00 &    0.06 \\
2703.39 &    0.07 &    2.23 &    0.42 &   -0.26 &    0.05 \\
2705.55 &    0.06 &    2.65 &    0.41 &   -0.44 &    0.05 \\
2708.68 &    0.06 &    2.90 &    0.51 &   -0.40 &    0.05 \\
2709.49 &    0.03 &    3.44 &    0.34 &    0.04 &    0.03 \\
2713.66 &    0.04 &    3.08 &    0.34 &    0.04 &    0.04 \\
2717.58 &    0.05 &    2.79 &    0.33 &   -0.31 &    0.04 \\
2720.75 &    0.03 &    2.70 &    0.26 &   -0.25 &    0.03 \\
2723.95 &    0.05 &    2.57 &    0.35 &   -0.48 &    0.05 \\
2727.92 &    0.05 &    2.66 &    0.38 &   -0.53 &    0.03 \\
2729.07 &    0.05 &    2.16 &    0.30 &   -0.30 &    0.04 \\
2732.37 &    0.03 &    2.35 &    0.27 &   -0.49 &    0.03 \\
2735.34 &    0.01 &    2.18 &    0.23 &   -0.36 &    0.02 \\
2737.18 &    0.03 &    2.05 &    0.22 &   -0.22 &    0.01 \\
2739.66 &    0.04 &    2.19 &    0.23 &   -0.27 &    0.02 \\
2741.63 &    0.03 &    2.23 &    0.22 &   -0.11 &    0.02 \\
2744.07 &    0.03 &    2.35 &    0.20 &    0.08 &    0.02 \\
2746.56 &    0.05 &    2.33 &    0.28 &    0.24 &    0.03 \\
2759.16 &    0.04 &    2.62 &    0.47 &    0.10 &    0.04 \\
2762.64 &    0.07 &    2.52 &    0.53 &    0.24 &    0.07 \\
2823.26 &    0.04 &    2.05 &    0.28 &    0.11 &    0.03 \\
2829.84 &    0.04 &    2.05 &    0.33 &   -0.75 &    0.04 \\
\tableline
\end{tabular}
\end{table}

\begin{table}
\tiny
\caption{Results from the global fitting program for systemic masers in UGC 3789.  The columns show (1) the                  
LSR velocity of the maser component at the reference epoch (2006 December 10) determined from the Gaussian fit, (2)
the uncertainty in the fit velocity, (3) the full width at half maximum of the Gaussian fit at the reference epoch,
(4) the uncertainty in the width, (5) the fit acceleration, (6) the uncertainty in the acceleration, and (7)
the number of epochs used in the global fit.  Velocities use the optical convention.  In this paper we model the 
systemic masers with emission from two thin rings.  The superscripts on the velocity entries in column 1 of this 
table mark those components that we identify either with ring 1 or ring 2.
}
\bigskip
\begin{tabular}{lrrrrrr}
\tableline\tableline
v$_{LSR}$ & $\sigma_{v}$ & Width & $\sigma_{width}$ & Accel & $\sigma_{accel}$ & No. epochs \\
(km s$^{-1})$ & (km s$^{-1}$) & (km s$^{-1}$) & (km s$^{-1})$ & (km s$^{-1}$ yr$^{-1}$) & (km s$^{-1}$ yr$^{-1}$) \\
\tableline
3233.33$^1$ &    0.04 &    2.63 &    0.33 &    1.97 &    0.06 & 30 \\
3238.87$^1$ &    0.04 &    2.94 &    0.31 &    1.22 &    0.04 & 30 \\
3243.70$^1$ &    0.03 &    3.00 &    0.24 &    1.38 &    0.03 & 30 \\
3248.18$^1$ &    0.02 &    2.54 &    0.22 &    1.29 &    0.02 & 15 \\
3252.55$^1$ &    0.04 &    2.52 &    0.31 &    0.51 &    0.04 & 15 \\
3256.13$^1$ &    0.05 &    2.17 &    0.26 &    1.44 &    0.12 & 15 \\
3258.70$^1$ &    0.06 &    2.05 &    0.26 &    1.37 &    0.13 & 15 \\
3260.75$^1$ &    0.05 &    1.97 &    0.25 &    1.68 &    0.10 & 15 \\
3263.40$^1$ &    0.03 &    1.99 &    0.19 &    1.91 &    0.04 & 15 \\
3267.18$^1$ &    0.02 &    2.28 &    0.22 &    1.55 &    0.10 & 15 \\
3269.88$^1$ &    0.22 &    2.00 &    0.60 &    2.14 &    1.79 & 11 \\
3271.01 &    0.06 &    1.96 &    0.46 &    3.90 &    1.03 & 11 \\
3272.76 &    0.03 &    2.12 &    0.31 &    7.29 &    0.22 & 11 \\
3282.42 &    0.12 &    2.04 &    0.33 &    1.47 &    0.65 & 11 \\
3284.68 &    0.13 &    2.03 &    0.32 &    2.16 &    0.66 & 11 \\
3286.87 &    0.14 &    2.02 &    0.32 &    1.61 &    0.59 & 11 \\
3288.65 &    0.08 &    2.01 &    0.30 &    1.58 &    0.43 & 11 \\
3290.70 &    0.10 &    2.03 &    0.28 &    1.84 &    0.39 & 11 \\
3292.65$^2$ &    0.07 &    2.11 &    0.25 &    4.26 &    0.27 & 11 \\
3294.86$^2$ &    0.07 &    2.00 &    0.20 &    5.83 &    0.31 & 11 \\
3296.03$^2$ &    0.12 &    2.22 &    0.26 &    5.09 &    0.77 & 11 \\
3298.26$^2$ &    0.04 &    2.24 &    0.19 &    3.56 &    0.32 & 11 \\
3300.98$^2$ &    0.03 &    2.14 &    0.14 &    4.00 &    0.18 & 11 \\
3303.26$^2$ &    0.05 &    2.06 &    0.19 &    4.29 &    0.29 & 11 \\
3305.41$^2$ &    0.06 &    2.09 &    0.20 &    7.07 &    0.32 & 11 \\
3307.77$^2$ &    0.06 &    2.06 &    0.20 &    5.57 &    0.25 & 11 \\
3309.72$^2$ &    0.05 &    2.08 &    0.17 &    4.64 &    0.27 & 11 \\
3312.08$^2$ &    0.05 &    2.08 &    0.20 &    3.80 &    0.25 & 11 \\
3315.01$^2$ &    0.08 &    2.05 &    0.27 &    4.42 &    0.39 & 11 \\
3317.17$^2$ &    0.07 &    2.03 &    0.26 &    7.02 &    0.45 & 11 \\
3319.50 &    0.12 &    2.01 &    0.30 &    5.98 &    0.92 & 11 \\
3321.58 &    0.24 &    2.00 &    0.31 &    4.49 &    1.59 & 11 \\
3324.70 &    0.25 &    1.99 &    0.31 &    4.32 &    1.36 & 11 \\
3326.37 &    0.28 &    2.01 &    0.31 &    6.57 &    2.30 & 11 \\
\tableline
\end{tabular}
\end{table}

\begin{table}
\tiny
\caption{Results from the global fitting program for ``red'' high-velocity features in UGC 3789.  The columns show (1) the                  
LSR velocity of the maser component at the reference epoch (2006 December 10) determined from the Gaussian fit, (2)
the uncertainty in the fit velocity, (3) the full width at half maximum of the Gaussian fit at the reference epoch,
(4) the uncertainty in the width, (5) the fit acceleration, and (6) the uncertainty in the acceleration.  Velocities
use the optical convention.}
\bigskip
\begin{tabular}{lrrrrr}
\tableline\tableline
v$_{LSR}$ & $\sigma_{v}$ & Width & $\sigma_{width}$ & Accel & $\sigma_{accel}$ \\
(km s$^{-1})$ & (km s$^{-1}$) & (km s$^{-1}$) & (km s$^{-1})$ & (km s$^{-1}$ yr$^{-1}$) & (km s$^{-1}$ yr$^{-1}$) \\
\tableline
3650.43 &    0.05 &    2.24 &    0.29 &   -0.16 &    0.03 \\
3653.24 &    0.05 &    2.27 &    0.29 &    0.05 &    0.04 \\
3656.86 &    0.04 &    2.52 &    0.24 &   -0.31 &    0.03 \\
3660.05 &    0.03 &    2.75 &    0.21 &   -0.01 &    0.02 \\
3663.82 &    0.04 &    2.57 &    0.28 &    0.11 &    0.03 \\
3668.81 &    0.05 &    2.49 &    0.28 &   -0.54 &    0.04 \\
3671.76 &    0.07 &    2.21 &    0.31 &   -0.41 &    0.05 \\
3676.88 &    0.07 &    2.22 &    0.32 &    0.23 &    0.05 \\
3680.86 &    0.08 &    2.17 &    0.32 &   -0.10 &    0.04 \\
3685.72 &    0.06 &    2.44 &    0.31 &    0.02 &    0.04 \\
3690.71 &    0.02 &    2.26 &    0.29 &   -0.56 &    0.04 \\
3694.31 &    0.07 &    2.29 &    0.32 &   -0.12 &    0.05 \\
3705.30 &    0.08 &    2.73 &    0.46 &   -0.56 &    0.04 \\
3714.22 &    0.06 &    2.95 &    0.45 &   -1.04 &    0.05 \\
3722.66 &    0.12 &    2.34 &    0.50 &   -0.33 &    0.10 \\
3732.18 &    0.05 &    2.54 &    0.27 &    0.05 &    0.03 \\
3734.87 &    0.04 &    2.22 &    0.25 &    0.13 &    0.03 \\
3737.12 &    0.04 &    2.23 &    0.20 &    0.09 &    0.02 \\
3739.61 &    0.03 &    2.06 &    0.19 &   -0.17 &    0.02 \\
3741.95 &    0.05 &    2.16 &    0.27 &   -0.26 &    0.03 \\
3744.56 &    0.04 &    2.25 &    0.26 &   -0.07 &    0.03 \\
3744.88 &    0.03 &    3.32 &    0.29 &   -0.29 &    0.03 \\
3748.43 &    0.06 &    2.64 &    0.31 &   -0.36 &    0.04 \\
3750.84 &    0.05 &    2.47 &    0.30 &   -0.08 &    0.03 \\
3754.00 &    0.04 &    2.49 &    0.28 &   -0.26 &    0.03 \\
3757.17 &    0.03 &    2.48 &    0.19 &   -0.61 &    0.02 \\
3759.68 &    0.05 &    2.48 &    0.26 &   -0.56 &    0.04 \\
3761.93 &    0.05 &    2.30 &    0.27 &   -0.19 &    0.03 \\
3764.35 &    0.06 &    2.36 &    0.33 &   -0.08 &    0.03 \\
3766.86 &    0.09 &    2.17 &    0.38 &    0.23 &    0.05 \\
3780.57 &    0.10 &    2.19 &    0.39 &    0.22 &    0.07 \\
3790.23 &    0.11 &    2.15 &    0.39 &   -0.53 &    0.09 \\
3807.35 &    0.04 &    2.60 &    0.35 &    0.98 &    0.03 \\
3814.04 &    0.08 &    2.55 &    0.42 &   -0.13 &    0.04 \\
3822.92 &    0.06 &    2.76 &    0.37 &    0.39 &    0.04 \\
3827.55 &    0.07 &    2.41 &    0.41 &   -0.37 &    0.06 \\
3831.19 &    0.06 &    2.59 &    0.38 &   -0.38 &    0.04 \\
3834.78 &    0.05 &    2.90 &    0.35 &   -0.12 &    0.04 \\
3838.63 &    0.03 &    3.23 &    0.28 &    0.24 &    0.02 \\
3842.77 &    0.04 &    3.05 &    0.31 &    0.91 &    0.02 \\
3847.96 &    0.04 &    3.01 &    0.37 &    0.12 &    0.04 \\
3852.82 &    0.05 &    3.12 &    0.37 &   -0.20 &    0.03 \\
3857.91 &    0.07 &    3.05 &    0.36 &   -0.28 &    0.05 \\
3861.09 &    0.04 &    3.17 &    0.26 &   -0.03 &    0.03 \\
3864.59 &    0.04 &    3.08 &    0.28 &    0.06 &    0.04 \\
3868.18 &    0.04 &    2.97 &    0.32 &   -0.34 &    0.04 \\
3869.33 &    0.04 &    3.21 &    0.36 &    0.10 &    0.03 \\
3874.14 &    0.04 &    3.07 &    0.35 &   -0.31 &    0.04 \\
3878.18 &    0.04 &    2.96 &    0.29 &   -0.47 &    0.03 \\
3881.27 &    0.04 &    3.05 &    0.24 &   -0.17 &    0.03 \\
3885.33 &    0.03 &    3.33 &    0.27 &   -0.22 &    0.03 \\
3892.63 &    0.06 &    4.10 &    0.47 &   -0.33 &    0.04 \\
3897.85 &    0.10 &    2.92 &    0.56 &    0.02 &    0.07 \\
3901.54 &    0.07 &    3.17 &    0.35 &   -0.28 &    0.05 \\
3904.55 &    0.05 &    2.83 &    0.32 &    0.03 &    0.04 \\
3908.18 &    0.05 &    2.97 &    0.30 &   -0.07 &    0.04 \\
3911.75 &    0.03 &    3.39 &    0.26 &   -0.01 &    0.03 \\
\tableline
\end{tabular}
\end{table}

\clearpage
\normalsize

\begin{figure}
\plotone{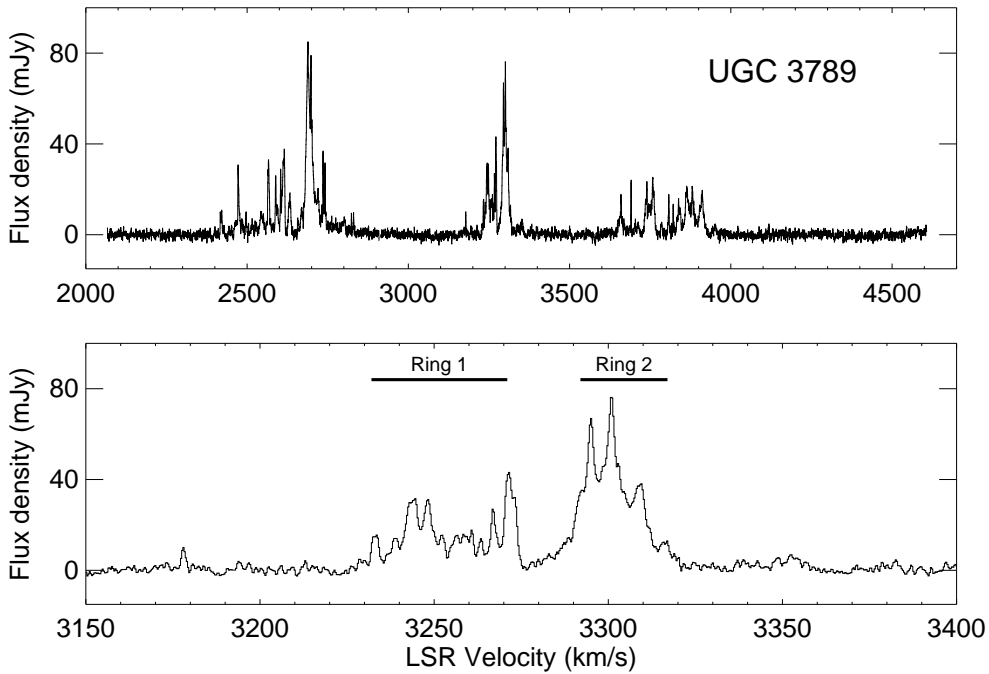}
\caption{A representative GBT spectrum of the water maser toward the nucleus of
UGC 3789.  This spectrum was observed on 2006 November 28, the monitoring epoch
just prior to the date of the VLBI observation, 2006 December 10.  The bottom panel 
shows the same spectrum zoomed in to show only the systemic features.
The velocity scale is referenced to the local standard of rest (LSR) and uses
the optical definition for velocity, defined as c$z$, where $z$ 
= $\Delta \lambda / \lambda$.  We model the systemic masers using two rings
(Section 5).  Horizontal bars in the lower panel mark the velocity ranges 
that we associate with each of the two rings.}
\end{figure}

\begin{figure}
\epsscale{.80}
\plotone{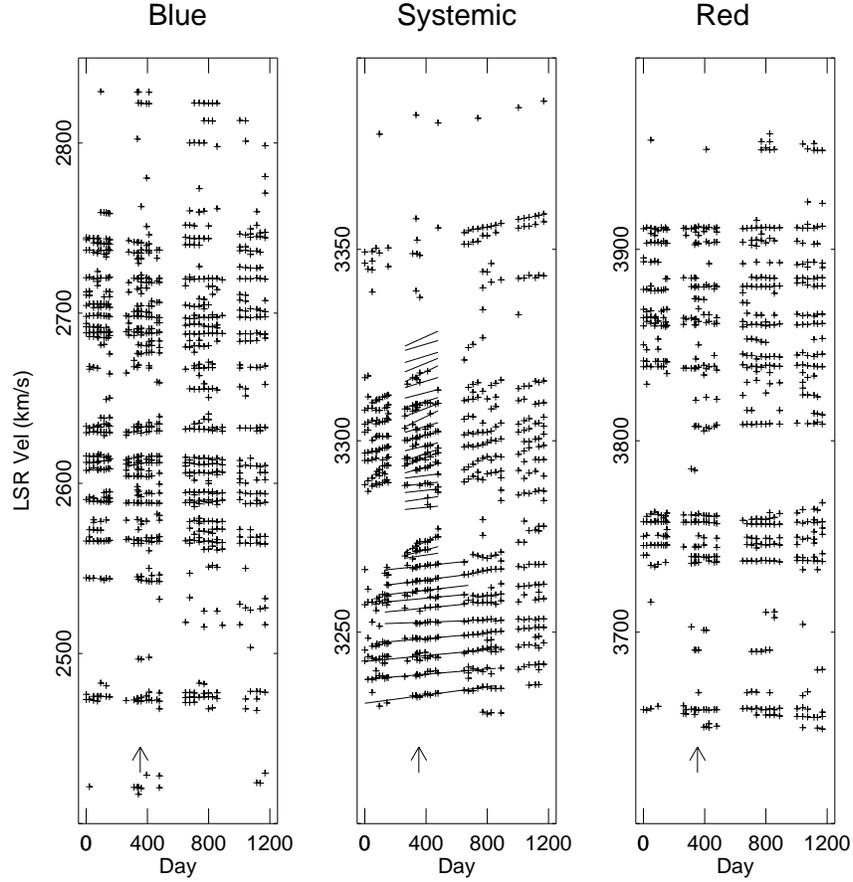}
\caption{The plus symbols show velocities of distinct maser peaks, determined by eye,
plotted against the epoch
observed.  The high-velocity blue and red features are plotted on the left and right 
panels, while the systemic masers are shown in the middle panel.  The positive slopes
evident in the systemic features reveal the centripetal acceleration of masers on the 
front side of the edge-on disk, while the flat trends in the high-velocity masers 
indicate they are located near the disk midline.  Day 0 is 2006 Jan 12.  The sloped line 
segments in the center panel represent the accelerations measured from the global fitting
of Gaussian components to the maser data (see Section 3).  The extent of each line segment indicates which 
epochs were used to fit that particular spectral component.  The fitting algorithm
is able to find stable solutions even in parts of the spectrum where there are no distinct
peaks marked in the spectral profile, as evidenced by solutions near 3320 km s$^{-1}$.
The arrow at the bottom of each panel indicates the date of the VLBI observation from Paper I.}
\end{figure}

\begin{figure}
\epsscale{.50}
\plotone{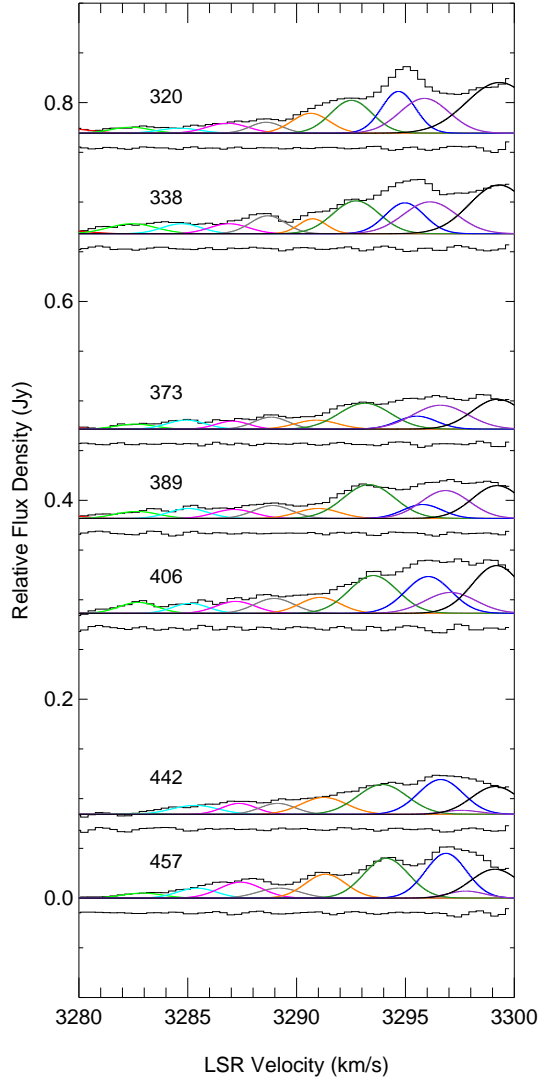}
\caption{An example of Gaussian decomposition for systemic maser features in UGC 3789.  
Data are shown for seven epochs.  The vertical offset for each epoch corresponds to the
date of the observation.  For each epoch, we show the original spectrum in the
upper histogram-style line.  The Gaussian components used to fit 
this data are plotted individually, each with a unique color so that common Gaussian
components can be identified from epoch to epoch.  The residual from each fit is also
shown, offset below the Gaussian components for clarity.  The number by each spectrum shows 
the day number for that epoch.}
\end{figure}

\begin{figure}
\epsscale{1.0}
\plotone{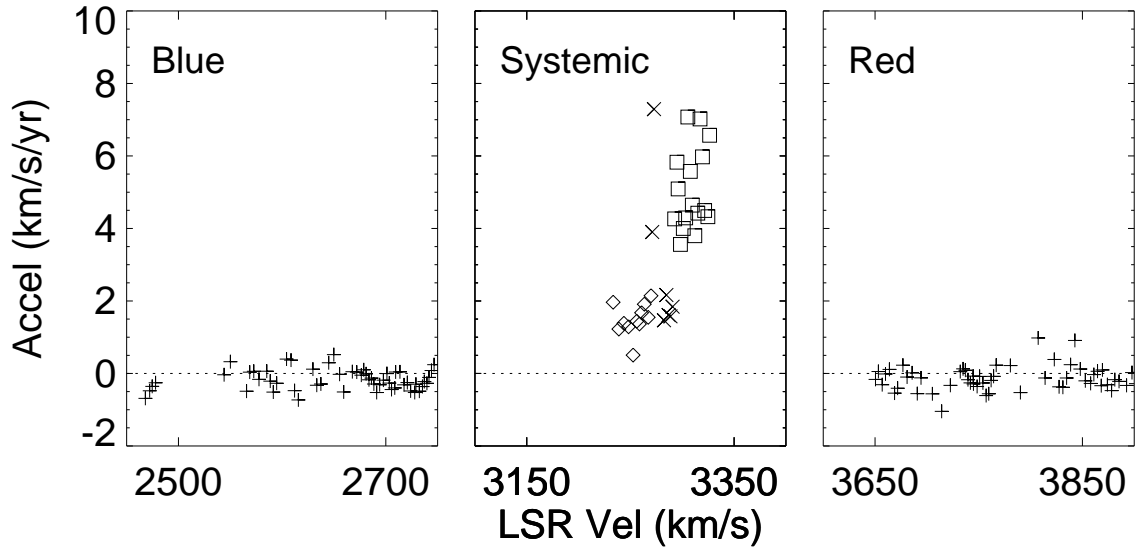}
\caption{Results of the global least-squares fits to measure centripetal acceleration 
of maser components from GBT monitoring data.  The velocities represented on the 
abscissa are from the reference epoch, 2006 December 10.  Each symbol on the plot
corresponds to an individual Gaussian maser component identified by the global
fitting program.  Maser components identified with ``ring 1'' are indicated with 
diamonds while those identified with ``ring 2'' are marked by squares.  Data points 
in the systemic range marked with crosses correspond to components in the velocity range not
identified with either ring, and were not directly used in the distance determination.}
\end{figure}

\begin{figure}
\plotone{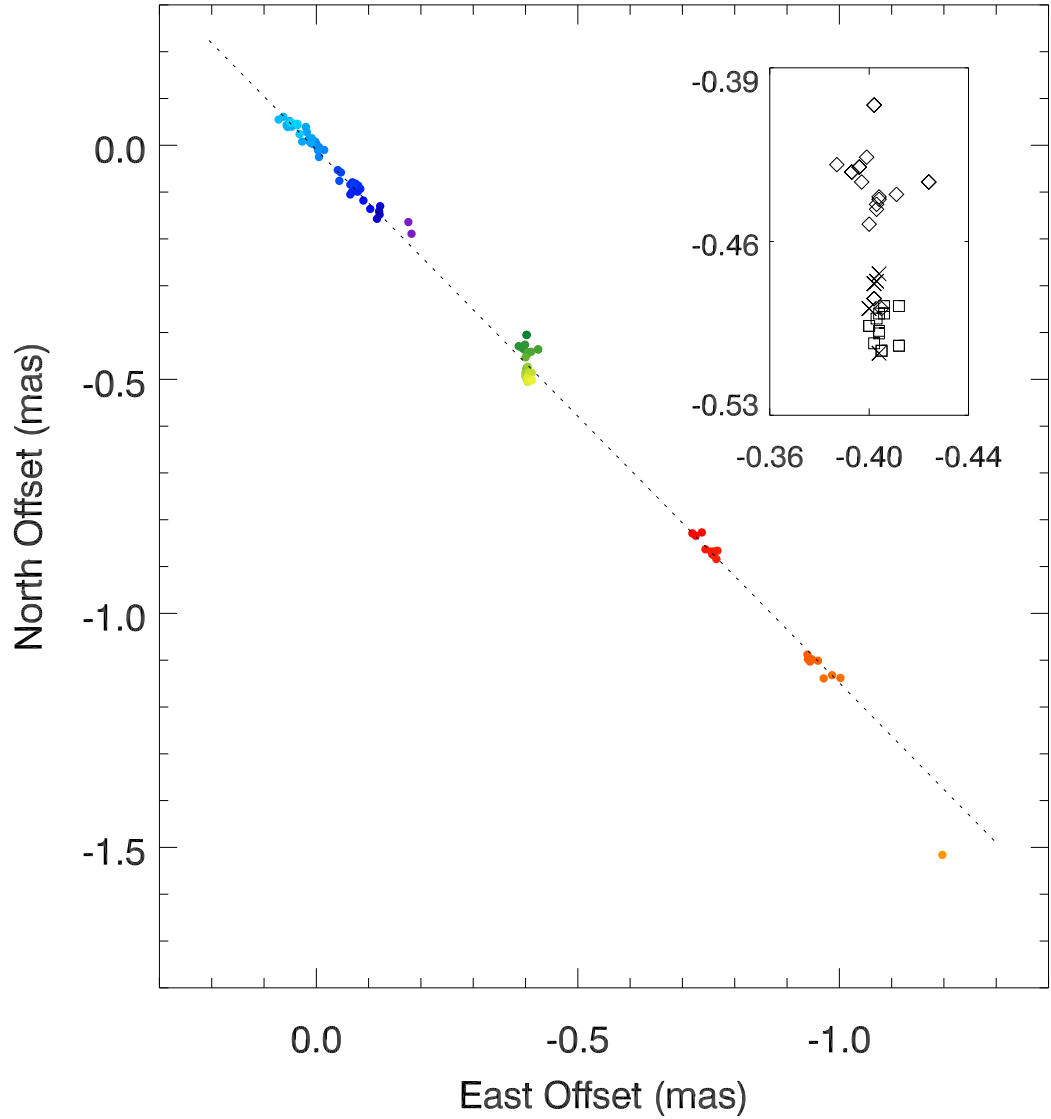}
\caption{A map of the relative positions of maser spots toward UGC 3789.  The 
representation in the large panel reproduces the map from Reid et al. (2009).
The inset zooms in to show only the systemic maser features.  Each symbol
represents the position of the maser emission from an individual spectral channel
in the VLBI map.  Masers associated with ring 1 are shown as diamonds and those with ring 2 as squares.}
\end{figure}

\begin{figure}
\plotone{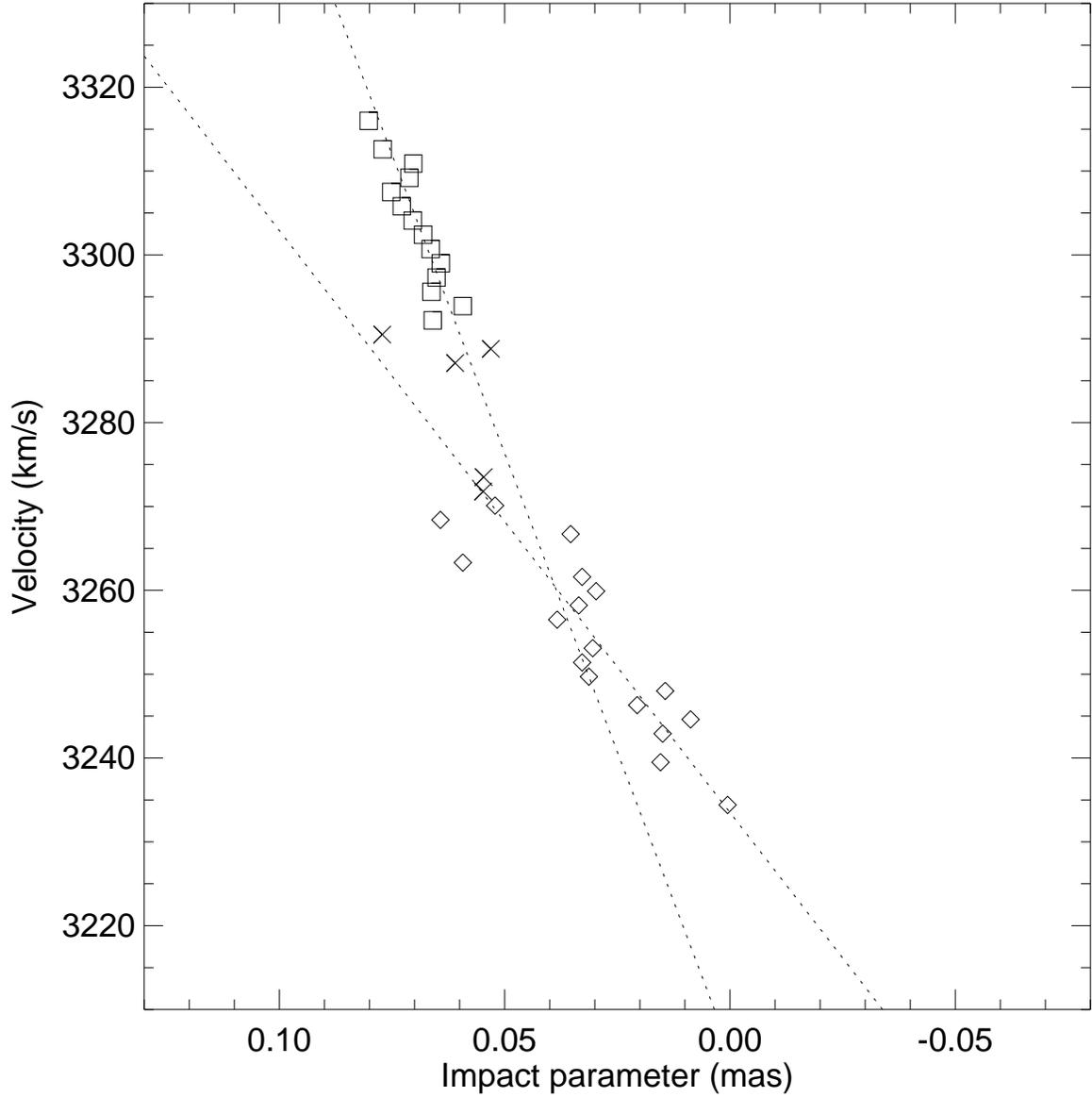}
\caption{The systemic maser features are plotted on a position-velocity diagram.  Each symbol
represents the position of the maser emission from an individual spectral channel in 
the VLBI map, as in Paper I.  The diamonds and squares show the maser features associated with 
rings 1 and 2, respectively.  The dotted lines show the linear fits to those features.}
\end{figure}

\end{document}